\documentclass[runningheads]{llncs}
\usepackage[english]{babel} 
\usepackage{graphicx}
\usepackage{algorithm}
\usepackage{algpseudocode}
\usepackage{amssymb}
\usepackage{amsmath}
\usepackage{makecell}
\begin{document}

\title{GAN-GA: A Generative Model based on Genetic Algorithm for Medical Image Generation}
\titlerunning{Medical Image Synthesis with GAN-GA}
%

\author{Mustafa AbdulRazek\inst{1, 3}\orcidID{0000-0002-5813-754X} \and
Ghada Khoriba\inst{1,2}\orcidID{0000-0001-7332-0759} \and
Mohammed Belal\inst{1}}
\authorrunning{M. AbdulRazek et al.}
%
\institute{Faculty of Computers and Artificial Intelligence, Helwan University, Cairo, Egypt 
\email{\{mustafa, ghada\_khoriba, belal\}@fci.helwan.edu.eg}
\and
Information Technology and Computer Science School, Nile University, Giza, Egypt
\email{ghadakhoriba@nu.edu.eg}
\and 
TachyHealth, Lewes, US \\
\email{mabdulrazek}@tachyhealth.com}
%
\maketitle              

\begin{abstract}
Medical imaging is an essential tool for diagnosing and treating diseases. However, lacking medical images can lead to inaccurate diagnoses and ineffective treatments. Generative models offer a promising solution for addressing medical image shortage problems due to their ability to generate new data from existing datasets and detect anomalies in this data.
Data augmentation with position augmentation methods like scaling, cropping, flipping, padding, rotation, and translation could lead to more overfitting in domains with little data, such as medical image data.
This paper proposes the GAN-GA, a generative model optimized by embedding a genetic algorithm. The proposed model enhances image fidelity and diversity while preserving distinctive features. The proposed  medical image synthesis approach improves the quality and fidelity of medical images, an essential aspect of image interpretation. To evaluate synthesized images: Frechet Inception Distance (FID) is used. The proposed GAN-GA model is tested by generating Acute lymphoblastic leukemia (ALL) medical images, an image dataset, and is the first time to be used in generative models. Our results were compared to those of InfoGAN as a baseline model. The experimental results show that the proposed optimized GAN-GA enhances FID scores by about 6.8\%, especially in earlier training epochs. The source code and dataset will be available at: https://github.com/Mustafa-AbdulRazek/InfoGAN-GA.

\keywords{{Generative Models \and InfoGAN \and Medical Image Generation \and Genetic Algorithms.}}
\end{abstract}

\section{Introduction}\label{related_work}
Several challenges arise when working with medical image datasets using deep learning approaches. The most known challenges are limited dataset size, imbalanced datasets, variability in image acquisition, annotation and labeling, and patient privacy and security. Medical image datasets are often smaller than others, making it challenging to train deep-learning models effectively. This is because deep learning models require a large amount of data to learn the features of the images accurately. Generating realistic medical images through generative adversarial networks  (GANs) can be used to train deep-learning models for medical image analysis, ultimately leading to more accurate and effective diagnoses and treatment plans. In addition, medical image datasets can often be imbalanced, with certain diseases or conditions being underrepresented in the dataset. This can lead to bias in the model and result in inadequate performance in detecting less common diseases or conditions. Deep learning frequently employs data augmentation in such data-limited circumstances to expand the data and avoid over-fitting.  Medical image synthesis is an active area of research, with applications in medical image analysis, simulation, and the training of machine learning models. Generative models are deep learning models that can learn the probability distribution of a dataset and generate new samples similar to the training data.

Generative models can create new data from scratch or comparable to already-existing data distribution. Generative models, such as variational autoencoders (VAEs) and GANs, have shown promising results in synthesizing medical images.  One of the main categories of techniques used to learn generative models from challenging real-world data is GAN \cite{GANs}. GANs (GAN and its derivatives) train a discriminator to discriminate between actual samples in the training dataset and fake samples created by the generator, in addition to employing a generator to create semantically meaningful data from conventional signal distributions. The generator creates samples that are more realistic than ever to trick the discriminator. The training process continues until the generator prevails in the adversarial game, or until the discriminator cannot determine whether a given sample is real or fake other than randomly. However, these models have shown promising results, and careful evaluation of the synthetic data is necessary to ensure its clinical relevance and quality. 

However, finding a Nash equilibrium of a non-convex game with continuous, high-dimensional parameters is necessary for training GANs.
Instead of learning the Nash equilibrium of a game, GANs are often taught using gradient descent methods intended to discover a low value of a cost function. These techniques might not converge when used to look for a Nash equilibrium. To boost GAN's performance, several measures have been included, including Kullback-Leibler divergence \cite{Nguyen}, absolute deviation \cite{zahooo}, and Wasserstein distance \cite{Arjovsky}.

The main contributions of this paper are:
\begin{itemize}

  \item Optimizing GANs for medical image synthesis using a genetic algorithm: Genetic algorithms use evolutionary principles, including crossover, mutation, and selection, to find the best solutions to a given problem. Integrating generative models with genetic algorithms produced a more exact and effective solution than baseline generative methods.
  \item Our proposed model was tested on Acute Lymphoblastic Leukemia, a small medical image dataset. The results show our model enhances generating performance as well as training stability. We create a generative adversarial network that addresses adversarial training as an evolutionary problem (GA-Embedded-with-InfoGAN). The embedding process of the genetic algorithm added another value by accelerating the learning and enhancement processes of model training.
\end{itemize}

\section{Related Work}\label{sec2}
Deep convolutional neural networks (CNN) are the most commonly used networks for image processing. Faster R-CNNs were proposed by Kang et al.\cite{Kang}; Zhang et al.\cite{inproceedings_Junkang} for the identification of urine particles and the detection of deep cancer cells, respectively. R-CNN has the advantage of extracting Region Proposal Network  (RPN) proposal areas that combine an image and produce a set of suggested bounding boxes with object ratings. Kang et al.\cite{Kang} utilized Faster R-CNN and SSD as state-of-the-art techniques for CNN-based object detection, along with numerous structural variants and the best AP for cast particles at 77.2 percent. Moreover, subsequent results demonstrate that, under conditions of limited cell adhesion, the sample had an accuracy of 0.979, a recall of 0.989, and an average precision of 0.908.

Additional research, including (Du et al.\cite{DU}; Xiaohui et al.\cite{8729545}), suggested a novel CNN design based on the analysis of principle components (PCA). Du et al. utilized R-CNN-based cell detecting methods consisting of two-part extraction algorithms, candidate recognition, and position regression. The algorithm has a detection accuracy of 93.6 percent and a detection time of 300 milliseconds per image. Recognizing a visual image of fecal composition, Xiaohui et al. achieved an accuracy of 90.7\% and a low time consumption (1200 ms) on photos of varying sizes.

Zhang et al.\cite{rs12071149}  offered an optimized GAN-based Hyperspectral classification model for a smooth training process and enhanced classification based on the gradient penalty for GAN's generating opponents (PG-GAN) and the Wasserstein generative network (WGAN-GP). They employed the PG-GAN training approach to make training fluid and the WGAN-GP loss feature to enhance convergence and balance in training. Their design considerably enhanced the GAN-based HSI classification approach.

To address gradient vanishing, divergence mismatching, and mode collapse, Zhaoyu et al. \cite{inproceedings_Zhaoyu}  developed a novel GAN consisting of one generator G and two discriminators (D1, D2) in the form of extensive CIFAR 10/100 and ImageNet dataset tests. First, the Spectral Standardization (SN) and ResBlock were applied in D1 and D2, focusing on the gradient vanishing. Half layers of D2 had incorporated Scaled Exponential Linear Units (SELU) as an additional solution to the problem.

Xin et al.\cite{review} presented a study of the recent development in medical imaging by implementing the adversarial training scheme that is very important in the visual community because of its ability to produce data without directly modeling the density of probabilities. In some instances, it has been proven helpful, such as adapting the domain, increasing the data, and converting images into an image. In addition to several positive GAN utilities, problems continue to be addressed in medical imaging. Most works also take conventional shallow benchmarks such as MAE, PSNR, or SSIM for quantitative evaluation in image restoration and cross-modality image synthesis. 

Liu et al.\cite{LIU202281} introduced an evolutionary algorithm-assisted GAN, dubbed EvoGAN, to produce multiple compound expressions with any precise goal compound expression. They transferred the compound expression synthesis into an optimization problem and employed an EA to search target results in the data distribution learned by the GAN. 
Liu et al. aimed to apply psychological studies or create compound expression databases. The outcomes of the experiments prove the viability and potential of EvoGAN. Furthermore, examinations were conducted using GANimation and VGG-19. The outcomes of the experiments prove the viability and potential of EvoGAN. \\
Modanwal et al. \cite{Modanwal} proposed  two solutions. The first one was to incorporate mutual information into the loss function. A method that performs intensity normalization and learns the noise distribution pattern. With improved accuracy, the proposed model can successfully learn a bidirectional mapping between MRIs produced by different vendors. The second solution to maintaining the breast's structure is modifying the discriminator. One limitation of their work is that it provides the capability of translation using 2D images only.
Anders et al. \cite{zombie} presented preliminary results showing that a 3D progressive growing GAN can synthesize MR brain volumes. They used T1-weighted MR volumes from the Human Connectome Project (HCP) for training  3D GAN. They performed data augmentation by applying ten random 3D rotations to each of the 900 volumes to achieve 9000 training volumes. They based their 3D progressive growing GAN (PGAN) on the 2D PGAN Tensorflow Implementation, replaced all 2D convolutions with 3D convolutions, and added an extra dimension to all relevant TensorFlow calls.

Ying et al. \cite{X2CT} presented a model to retract CT from the two orthogonal X-rays with the GAN framework to increase 2D (X-rays) to 3D data dimension (CT).
They mixed the loss of reconstruction, the loss of projection, and the loss in the GAN. Qualitatively and quantitatively, studies have shown that biplanar X-rays
in 3D reconstruction, techniques are superior to single-visual X-rays.

Geng et al. \cite{FFusionCGAN} developed a method for generating fused images based on conditional GANs from one- or few-focus images. Geng et al. method can generate fused images with transparent textures and deep field depths.
The model was developed to learn to map input source images to fused images directly, without the need to manually construct complex measurements of activity level and fusion rules in conventional ways. Unattended model expansion to various datasets was not yet applied.

Chen et al.\cite{Chen_infoGAN} proposed Information Maximizing Generative Adversarial Networks (InfoGAN), an unsupervised representation learning technique that learns interpretable and disentangled representations on many datasets. One of their objectives was to investigate whether changing a specific latent factor results in only one form of semantic variance in the generated images. The second objective was to establish if InfoGAN can discover disentangled and interpretable representations (that are competitive with representations learned by existing supervised methods).

Carbonne et al. \cite{7529319} proposed and developed a machine
learning algorithm based on a genetic algorithm for profile recognition. 
Their solution combines natural language processing, evolutionary algorithm, and supervised machine learning. They recalled how to represent a profile in a Vector Space Model in order to ease the processing of semantic data. Then, this model goes through a Genetic Algorithm (GA) used as a supervised learning algorithm for training a computer to determine how similar two profiles are. The principle of genetic algorithms is used to train the computer to evaluate the significance of each label. The model trained with the GA over a training set composed of 3,003 similarities between profiles. In the training set, an output for the similarity between two profiles is either 1 (the two profiles correspond to the same person) or 0. They tested the model predictions within an entirely new dataset.

While deep learning models such as GANs and VAEs have demonstrated promising outcomes, several obstacles still need to be overcome to improve synthetic image quality while keeping distinctive features.
We propose a Generative Model embedded with Genetic Algorithm (InfoGAN-GA) to address these issues for diverse medical image generation. Our research intends to develop a method for enhancing the quality and diversity of generated images with fewer computational resources.

\section{Proposed InfoGAN-GA: Generative Model embedded with Genetic Algorithm}
Although deep learning-based methods have produced promising outcomes, they are occasionally limited by a lack of diversity in generated images, which can lead to biased or misleading diagnoses.
We propose an InfoGAN-GA model that integrates a genetic algorithm within the Information Maximizing GAN (InfoGAN) architecture originally introduced by Chen et al. \cite{Chen_infoGAN}. This integration enables the InfoGAN-GA to leverage the strengths of the InfoGAN and the genetic algorithm, resulting in an advanced generative model that performs betters on image generation tasks in terms of fidelity and converges in fewer epochs than vanilla InfoGAN.
In this section, we first review the InfoGAN and Genetic Algorithm formulation. Then, we introduce the proposed Generative Model embedded with Genetic Algorithm InfoGAN-GA.

\subsection{Generative Adversarial Networks (GANs)}
\begin{figure}
 \centering
 \includegraphics[width=\textwidth,height=\textheight,keepaspectratio]{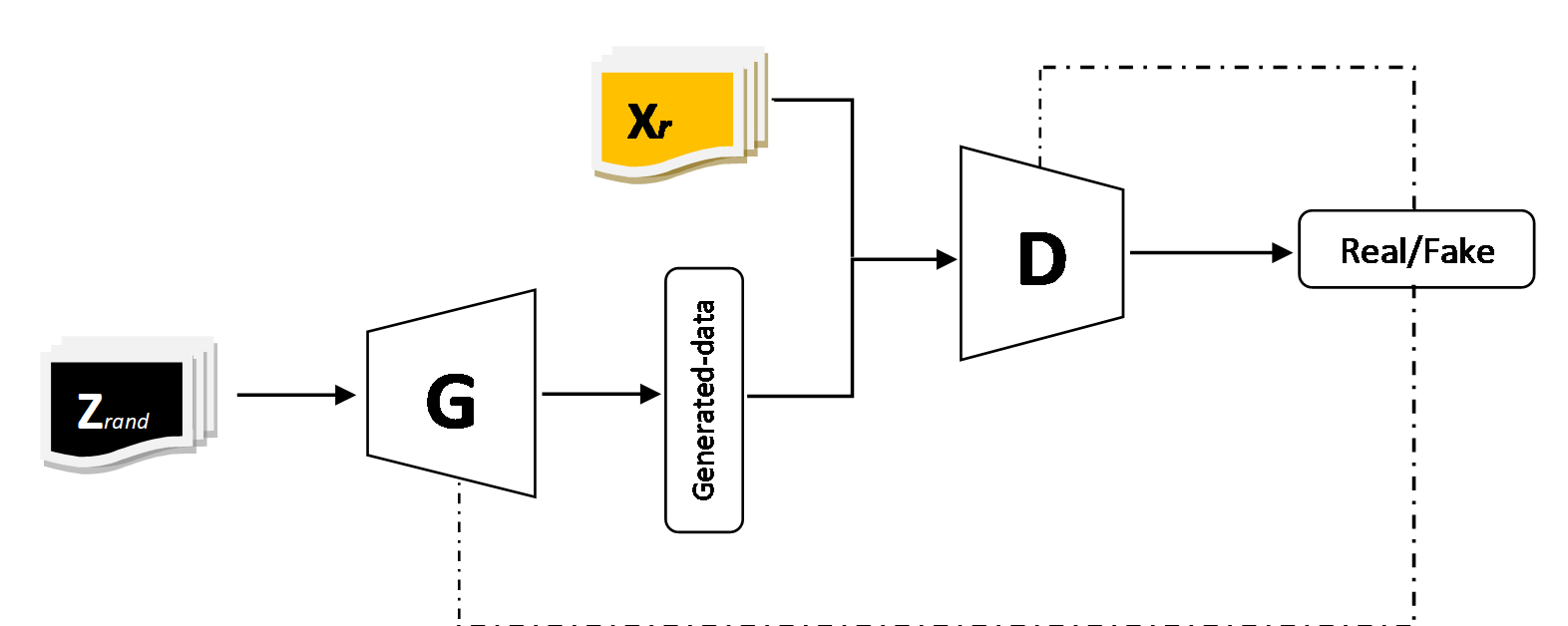} 
 \caption{Original Generative Adversarial Nets (GAN)} \label{fig-original-GAN}
\end{figure}

A two-player mini-max game between a discriminator network $D$ and a generative network $G$ is studied by GAN, which was first proposed in \cite{GANs}. The generative network $G$ outputs new data $G(z)$, whose distribution $p_g$ is meant to be similar to that of the data distribution $p_{data}$, given the noisy sample $z \sim p(z)$ (sampled from a uniform or normal distribution) as the input. In the meantime, the generated sample and the ground-truth data sample $p_{data}(x)$ are separated using the discriminator network $D$.  
In the original GAN, the adversarial training process was formulated as follows:
\begin{equation}
\min_{G}\max_{D}\mathbb{E}_{x\sim p_{\text{data}}(x)}[\log{D(x)}]+\mathbb{E}_{z\sim p_{z}(z)}[1-\log{D(G(z))}] 
\end{equation}
 The adversarial procedure is illustrated in Fig. \ref{fig-original-GAN}. Most existing GANs perform a similar adversarial procedure in different adversarial objective functions.
\subsection{Information Maximizing Generative Adversarial Nets (InfoGAN)}\label{InfoGAN}
InfoGAN, originally proposed in \cite{Chen_infoGAN}, can learn interpretable representations that are competitive with those acquired by existing used supervised learning techniques. Instead of using a single unstructured noise vector, InfoGAN decomposes the input noise vector into two parts: (i) $z$, which is treated as a source of incompressible noise; and (ii) $c$, the latent code that targets the salient structured semantic features of the data distribution. InfoGAN can learn disentangled representations entirely unsupervised manner and maximize the mutual information. A further term, $-\lambda{I(c; g(z,c))}$, separates InfoGAN's loss functions from GANs. $\lambda$ is a tiny positive constant.
Maximizing the mutual information $I(c;g(z,c))$ and minimizing the loss function of an InfoGAN (equations: \ref{eq:2} and \ref{eq:3}) equate to minimizing the loss of the original GAN.
\begin{equation}\label{eq:2}
L_D = -\mathbb{E}_{x\sim p_{\text{data}}}\log{D(x)} - \mathbb{E}_{z,c}\log{[1-D(g(z,c))]} - \lambda I(c;g(z,c))
\end{equation}
\begin{equation}\label{eq:3}
L_G = -\mathbb{E}_{z,c}\log{D(g(z,c))} - \lambda I(c;g(z,c))
\end{equation} 

\subsection{Genetic Algorithm}
In a genetic algorithm (GA) \cite{russel2010}, successor states are generated by combining two parent states instead of modifying a single state. This is an example of stochastic beam search variation.
Each state or individual is represented as a string across a finite alphabet and ranked by the objective  or fitness function. The likelihood of being picked for reproduction is related to the fitness score, and the corresponding percentages are displayed alongside the raw numbers. GAs begin with a population, which consists of k randomly generated states.
A random crossover point is chosen from the string positions to couple each crossover pair.
At the crossover point, the parent strings are crossed over to generate the offspring, which can result in a state that is significantly distinct from either parent state.
At the early stages of the search phase, when the population is often very diverse, crossover often occurs in large steps within the state space, followed by smaller steps when the bulk of individuals are similar.
The resulting children are then introduced to a new population, and the selection, crossover, and mutation processes are repeated until the new population is complete. Lastly, each position has a small possibility of being subject to random mutation independently.
Genetic algorithms incorporate an incline, random exploration, and information sharing across parallel search threads.
Genetic algorithms are primarily characterized by their crossover operation.
Intuitively, the effect results from crossover's ability to combine large chunks that have independently evolved to do meaningful tasks, enhancing the search's granularity.

\subsection{InfoGAN-GA: The Proposed Approach}
\begin{figure}
 \includegraphics[width=\textwidth,height=\textheight,keepaspectratio]{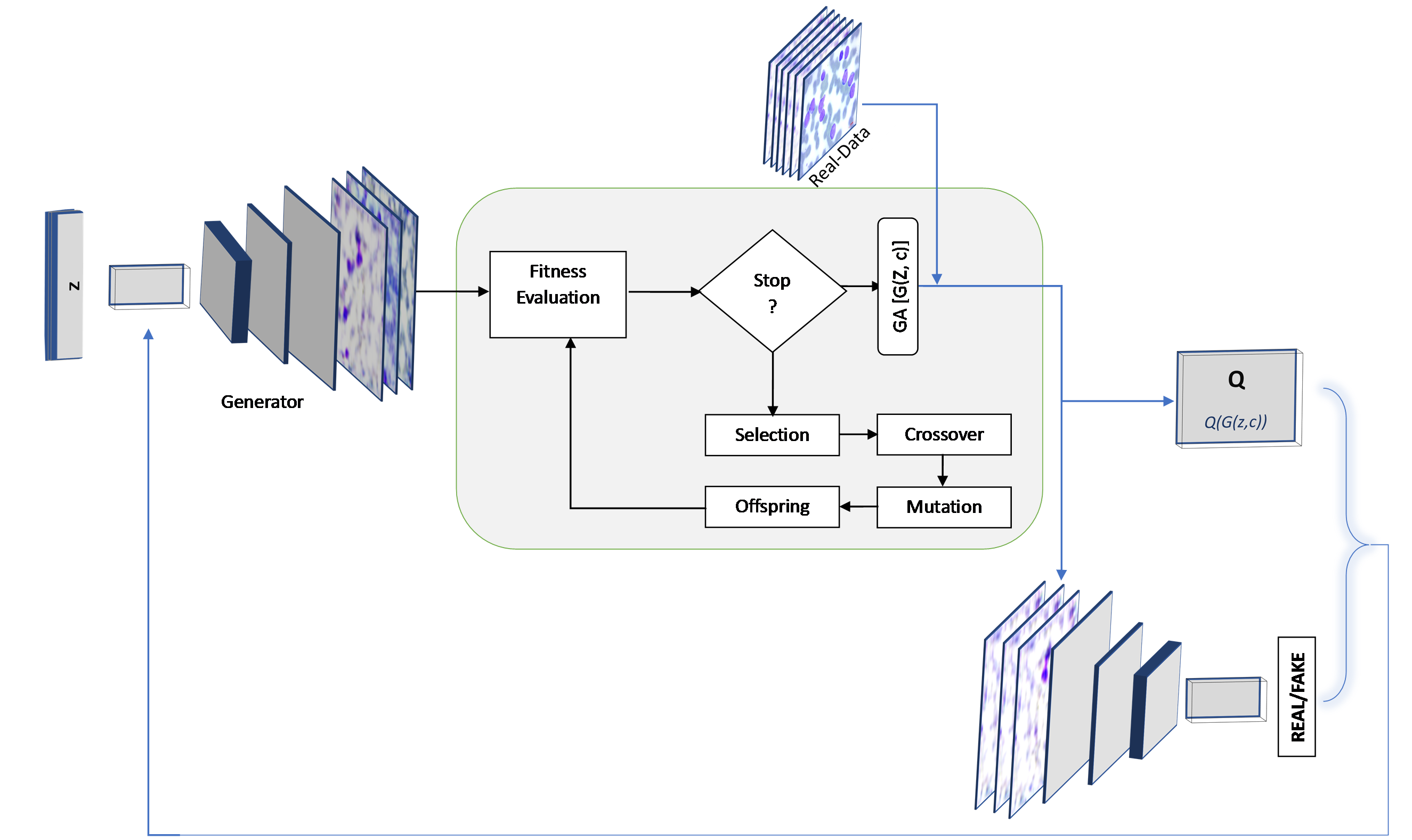} 
 \caption{Generative Model embedded
with Genetic Algorithm (InfoGAN-GA) Architecture. Latent vector z as input to Generator. $GA[G(Z, c)]$ is the output of genetic algorithm.  Q is an auxiliary network attached to the second to last layer of the discriminator inspired from original InfoGAN.} \label{fig-infoGAN-GA}
\end{figure}

An InfoGAN embedded with Genetic Algorithm (alg.\ref{algo1}) is proposed as shown in figure \ref{fig-infoGAN-GA}. By applying the GA to each training phase, we aimed to improve the generator's performance and speed up the generation of diverse images (as fake for the GAN) in earlier epochs. 
The initial population was first populated with the fake individuals $G(z)$ (chromosome) generated by the $G$, and the fitness of each fake individual was determined as the D's discrimination value $D(G(z))$. The selection type, a roulette wheel selection, and the arithmetic recombination crossover type were employed. The mutation was created by randomly adding a number to each gene with a 10\% probability of doing so. The population was evolved using the aforementioned parameters utilizing the high-fitness samples (recognized as real by the Discriminator).
The least fit member of the conventional population was replaced by the child when the fitness of the offspring was higher than that of the parent.

\algrenewcommand\algorithmicrequire{\textbf{Input:}}
\algrenewcommand\algorithmicensure{\textbf{Output:}}
\begin{algorithm}
\caption{Summary of InfoGAN-GA model}\label{algo1}
\begin{algorithmic}[1]
\Require {noise latent vector (z), real images (x), G and D}
\Ensure {synthesized images (Generated), FID score, G-loss and D-loss} 

\For{\texttt{<n of training iterations>}}
        \State \texttt{<Random Sample a batch of ($z\footnotemark[1]$ ... $z^m$) $p_{\text{g}}(z)$ >}
        \State \texttt{<Random Sample a batch of ($x\footnotemark[1]$ ... $x^m$) $p_{\text{data}}(x)$ >}
        \State $population \Leftarrow G(z) $
        \For{\texttt{<k in generations>}}
            \State \texttt{<Genetic Algorithm Optimization - Selection, Crossover, Mutation>}
            \State \texttt{<select best synthesized images>}
            \State \texttt{<D evaluation>}
        \EndFor
        \State \texttt{<update G, D>}
\EndFor
\end{algorithmic}
\end{algorithm}
\bigskip



\section{Results and Discussions}
\subsection{Experimental Settings}
\begin{figure}
 \centering
 \includegraphics[width=.9\columnwidth]{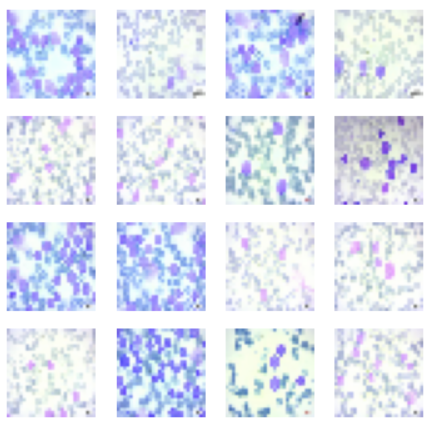} 
 \caption{Acute lymphoblastic leukemia (ALL) image dataset sample images 28x28.} \label{fig-samples}
\end{figure}

\hspace{3ex}Aria et al.\cite{Mehrad_Aria_Mustafa} created Acute lymphoblastic leukaemia (ALL) [table~\ref{table-data}] dataset and made available to the public. The bone marrow laboratory at Taleqani Hospital created the images for this dataset (Tehran). This dataset comprised 3256 PBS pictures from 89 patients (samples shown in fig.~\ref{fig-samples}) who were thought to have ALL and whose blood samples were skillfully processed and stained by laboratory personnel. The classes of benign and malignant data are separated in this data collection. The former includes hematogenous that are quite similar to ALL instances, but this hematopoietic precursor cell is benign, doesn't require chemotherapy, and typically goes away on its own. The latter category includes the malignant lymphoblasts known as (ALL) and the early pre-B, pre-B ALL, and pro-B ALL subtypes. A Zeiss camera set at 100x magnification was used to capture each image, which was then saved as a jpg file.
\begin{table}
 \centering
  \caption{Acute Lymphoblastic Leukemia (ALL) Image Dataset Description} \label{table-data}
 {\small
 \resizebox{0.9\width}{!}
 {
 \begin{tabular}{|c|c|c|c|}
  \hline
  \thead{Type} & \thead{Subtype} & \thead{No. of samples} & \thead{No. patients} \\
  \hline
  
  Benign    & Hematogones & $504$ & $25$ \\
  Malignant & Early pre-B ALL & $985$ & $20$ \\
            & Pre-B ALL & $963$ & $21$  \\
            & Pro-B ALL & $804$ & $23$  \\
            & Total & $3256$ & $89$   \\
  \hline
 \end{tabular}
}}
\end{table}

\subsection{Evaluation Results}
\begin{figure}[H] 
 \centering
 \includegraphics[width=\textwidth,height=\textheight,keepaspectratio]{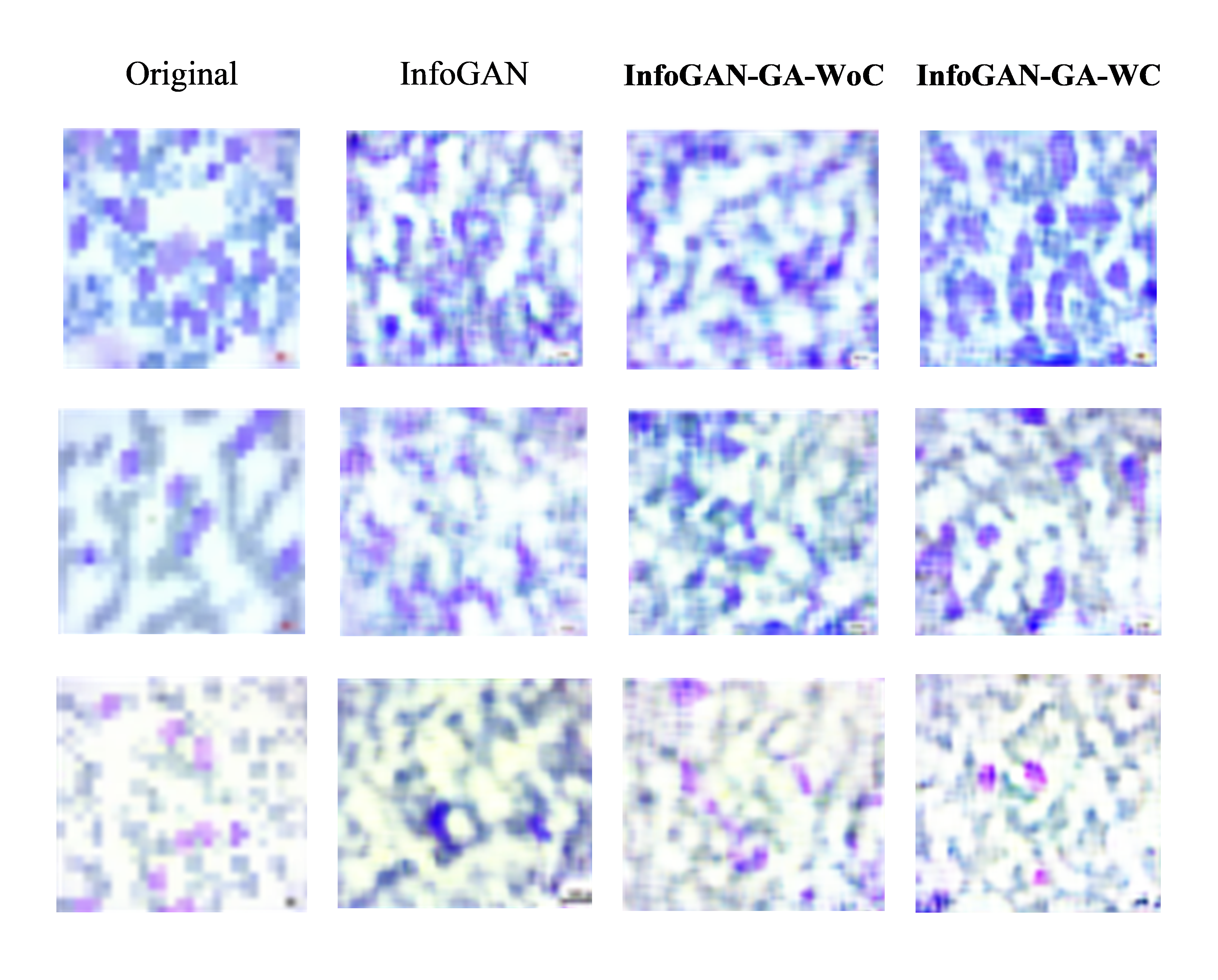} 
 \caption{Illustration of Original vs. Generated images from InfoGAN and InfoGAN-GA versions 28x28.} \label{fig-compare-results}
\end{figure}

\hspace{3ex}Assessing the effectiveness of Generative Adversarial Networks (GANs) is a challenging task, given the rapid advancements in this active research field. The measures used to evaluate GAN models apply to models other than ours, allowing for independent evaluations. GANs necessitate the production of images from random noise, in contrast to supervised learning with a classifier, where tests with labels are available for image categorization. It is impossible to evaluate the quality of these images because there is no clear objective for creating certain pixels depending on the input noise vector. The generated images are evaluated based on their realism or fidelity to generate various images. Discriminators in GANs are incapable of achieving perfection and frequently overfit to distinguishing real from false images for their particular generator, making it difficult to compare and rank different models. Lately, the most popular assessment metrics for GANs have been the Inception Score (IS) and Fréchet Inception Distance (FID), both of which rely on an ImageNet-trained classifier (InceptionNet).
Salimans et al.\cite{NIPS2016_8a3363ab} compute the KL divergence between the conditional class distribution and the marginal class distribution over the generated data. FID (Heusel et al.\cite{HeuselRUNKH17}) calculates the Wasserstein-2 distance between multivariate Gaussians fitted to the embedding space of the Inception-v3 network of generated and real images proposed as an improvement to the Inception Score (IS) \cite{NIPS2016_8a3363ab}. \\
We evaluated our model on "ALL" dataset \cite{Mehrad_Aria_Mustafa} at 28x28 resolutions. It's the first time to use like dataset with Generative Models, and the big challenge with the obtained dataset is the data size. The lack of sufficiently large datasets in the medical sector is one of the most common problems when dealing with medical AI solutions using ML and Deep learning models. This is because the clinical data is usually private, and the hospitals rarely release datasets to the general public. To overcome this, researchers initially considered image augmentation and up-scaling the non-tumorous scans. However, the problem is that the augmented images aren't very diverse, and models need new scans to train efficiently. The samples generated from our model are presented in Figure~\ref{fig-compare-results}, columns 3 and 4. 

\newcommand{\bftab}{\fontseries{b}\selectfont}
\begin{table}[]
 \centering
  \caption{Final Results and Relative Error Reduction} \label{table-results}
 {
 \resizebox{.9\columnwidth}{!}
 {\begin{tabular}{|c|c|c|c|c|c|}
  \hline
  \thead{Model}  & \thead{G Loss} & \thead{D Loss} & \thead{FID} & \thead{Starting FID} & \thead{CE}  \\
  \hline
  
  InfoGAN & $0.195848$ & $0.218547$ & $141$ & $328$ & $2700$  \\
 \bftab InfoGAN-GA-WoC & \bftab 0.248143 & \bftab 0.221726 & \bftab 127 & \bftab 226 & \bftab 2500   \\
 \bftab InfoGAN-GA-WC & \bftab 0.091876 & \bftab 0.224233 & \bftab 123 & \bftab 223  & \bftab 1750   \\
\hline
\end{tabular}}}
\end{table}

We show FID in Table~\ref{table-results}. The network architectures are based on InfoGAN \cite{Chen_infoGAN} and are briefly introduced here. We use the default
hyper-parameter values listed in default InfoGAN \cite{Chen_infoGAN} for all experiments. Our experiments show that InfoGAN-GA already outperforms 
InfoGAN model. Furthermore, all experiments were trained on NVIDIA GeForce RTX 2070 with Max-Q Design GPU and input/output 28×28 images.\\

Table~\ref{table-results} lists trained models, generator losses, discriminator losses, FID, starting FID, and convergence epoch (CE) values, respectively. It's obvious that the proposed model outperforms the original 
InfoGAN in final with FID score=123 compared with FID scores= 141, respectively.  

\begin{figure}
 \centering
 \includegraphics[width=\textwidth,height=\textheight,keepaspectratio]{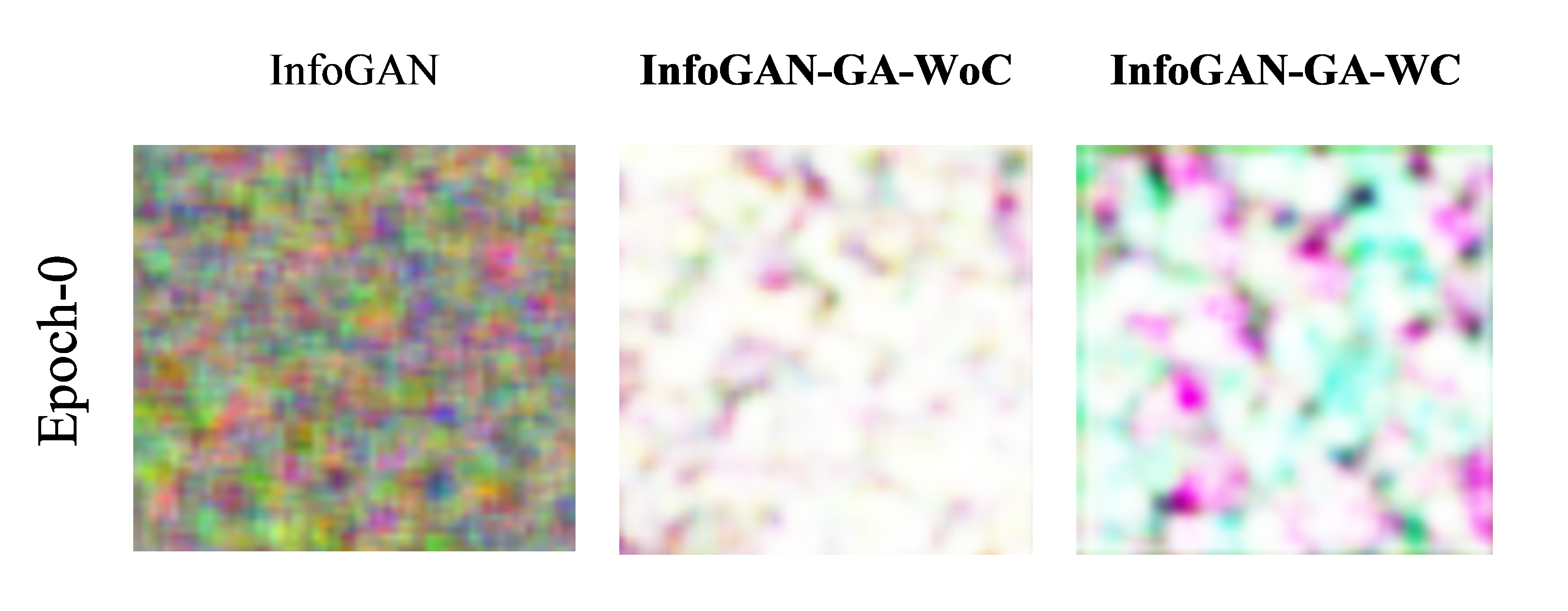} 
 \caption{Comparison between synthesized images after first epoch for base InfoGAN, InfoGAN-GA-WoC, and InfoGAN-GA-WC models.} \label{epoch-0}
\end{figure}

\begin{figure} 
 \centering
 \includegraphics[width=0.9\textwidth, height=5.2cm]{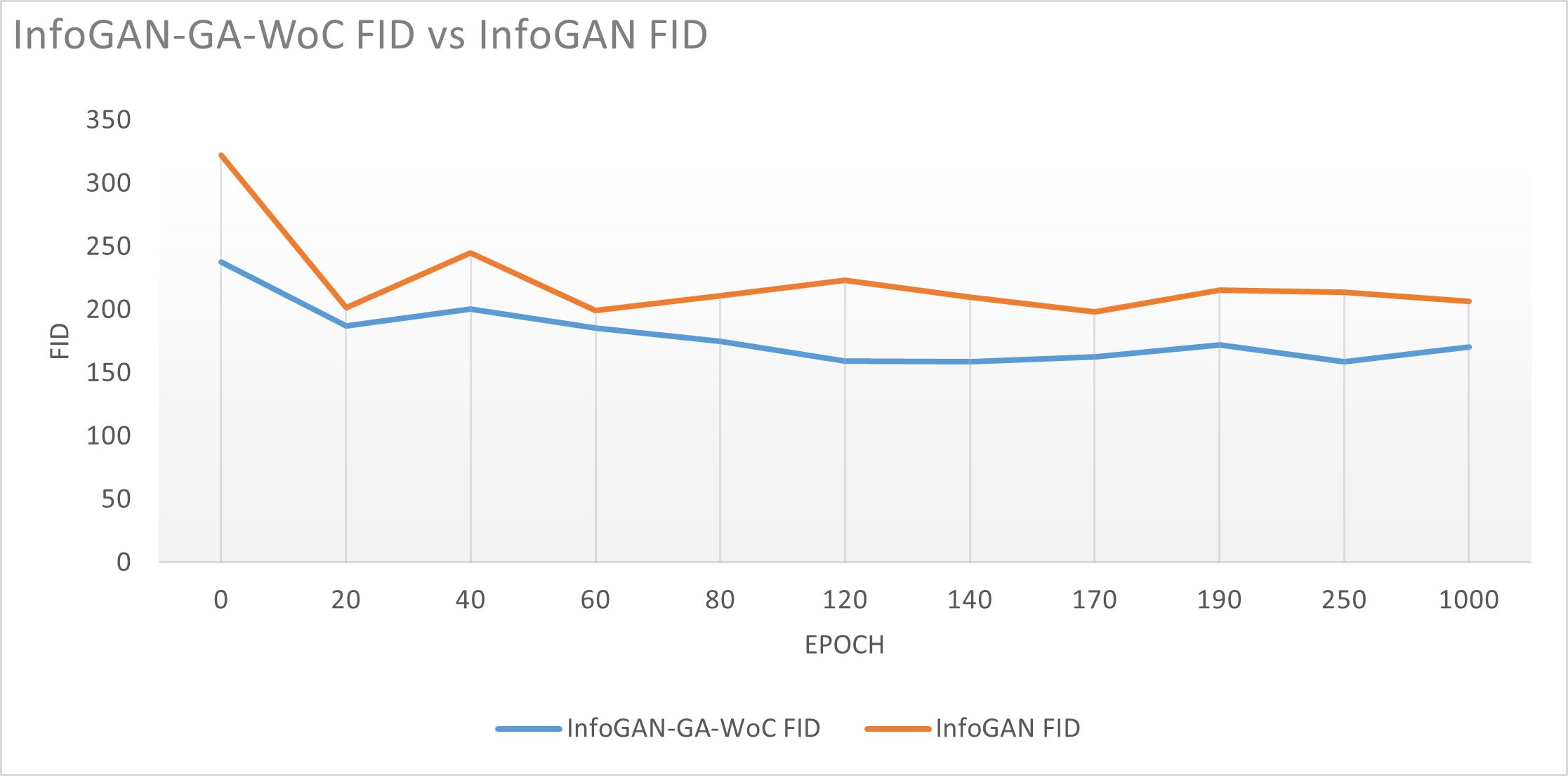} 
 \caption{Comparison for InfoGAN vs. InfoGAN-GA-WoC using FID Scores through epochs.} \label{fig-fid-lower}
\end{figure}

\begin{figure} 
 \centering
 \includegraphics[width=0.9\textwidth, height=5.2cm]{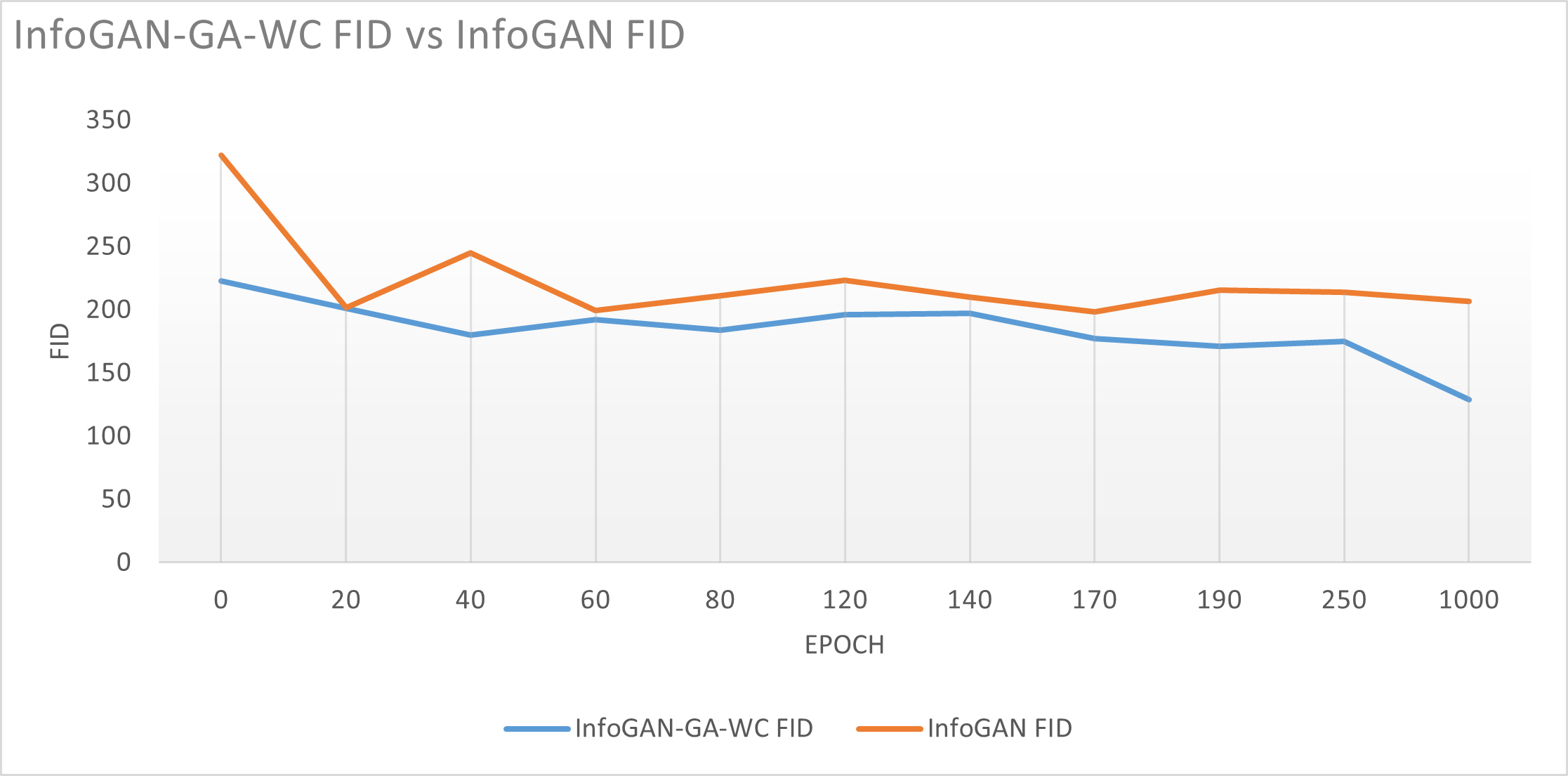} 
 \caption{Comparison for InfoGAN vs InfoGAN-GA-WC using FID Scores through epochs.} \label{fig-fid-lower-v02}
\end{figure}

 The embedding process of the Genetic Algorithm accelerated enhancement of generated images quality (as proofed in figure~\ref{epoch-0}, synthesized images after the first epoch, it's evident that synthesized images in earlier epochs in the proposed model are better quality than InfoGAN base model) faster than other mentioned previous base models as shown in figure \ref{fig-fid-lower}, \ref{fig-fid-lower-v02} and presented starting FID score in table~\ref{table-results}, and achieved the diversity through feeding only the discriminator with real images but feeding the Genetic Algorithm with population generated through Generator and removed the process of feeding the Genetic with target images not to generate identical images.  

The embedding process of the Genetic Algorithm accelerated the enhancement of generated image quality. As shown in figure~\ref{epoch-0}, Synthesized images after the first epoch prove synthesized images in earlier epochs in the proposed model are of higher quality than the InfoGAN base model as shown in the figure and presented starting FID score in the table. Proposed InfoGAN-GA achieved diversity by feeding only the discriminator with real images but feeding the Gene Pool of genetic algorithm with synthetic images.

As shown in Figure~\ref{fig-fid-lower}, \ref{fig-fid-lower-v02}, the effect of adding Genetic Algorithm optimization methodology between the generator and discriminator is obvious here in optimizing FID scores, especially in early epochs. The starting FID for the first epoch of the proposed model is better than the ones of InfoGAN (base model). Due to the Genetic Algorithm's effect of picking the best population formed by the generator model and feeding just the best-selected individuals to the discriminator, as stated previously.

\begin{figure} 
 \centering
 \includegraphics[width=0.9\columnwidth, height=4.0cm]{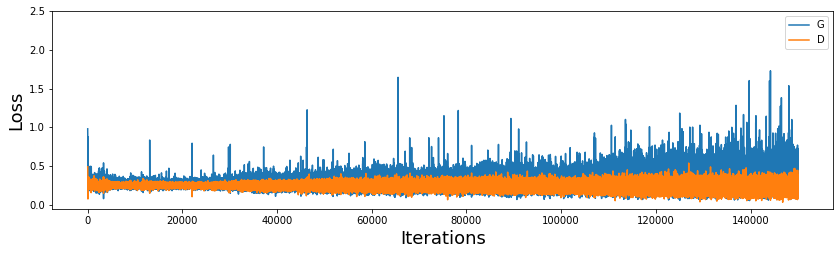} 
 \caption{Generator and Discriminator Loss During Training (InfoGAN) 3000 epochs.} \label{fig-info-loss}
\end{figure}

\begin{figure} 
 \centering
 \includegraphics[width=0.9\columnwidth, height=4.0cm]{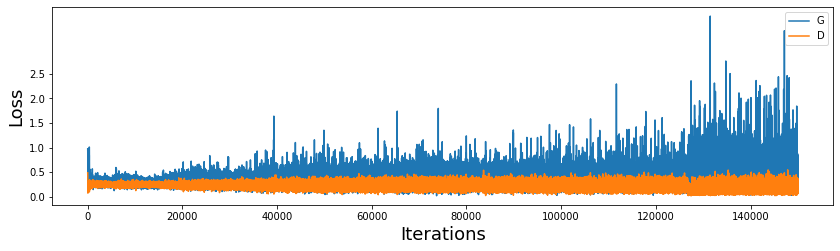} 
 \caption{Generator and Discriminator Loss During Training (InfoGAN-GA-WoC) 3000 epochs.} \label{fig-GA-info-loss}
\end{figure}

\begin{figure} 
 \centering
 \includegraphics[width=0.9\columnwidth, height=4.0cm]{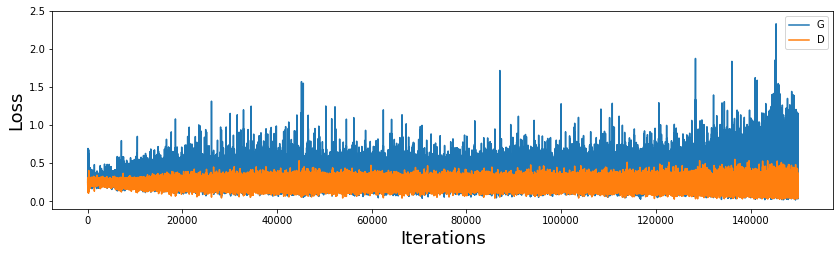}
 \caption{Generator and Discriminator Loss During Training (InfoGAN-GA-WC) 3000 epochs.} \label{fig-GA-info-loss-v02}
\end{figure}

The losses of the generator and discriminator during training for each InfoGAN algorithm and the proposed algorithm (InfoGAN-GA versions) are shown in Figures~\ref{fig-info-loss}, ~\ref{fig-GA-info-loss}, ~\ref{fig-GA-info-loss-v02} respectively.

According to the original GAN citation, \cite{GANs}, the generator and discriminator are constantly competitive, with gains in one leading to losses in the other until the latter learns to exploit the loss of information acquired from the former. In most cases (depending on data and initialization), both discriminator and generator losses should converge to a fixed number. This loss convergence would normally indicate that the GAN model has reached an optimal point where it cannot improve further, which should also indicate that it has learned well enough.

From figurer~\ref{fig-GA-info-loss} and \ref{fig-GA-info-loss-v02}, the generator's performance improves with subsequent epochs, whereas the discriminator's performance deteriorates when the discriminator settles to a value. It is evident that generator losses result in unstable alterations that improve new (false) image generation. The convergence speed and stability during the training process were enhanced compared to the GAN base model.

 \section{Conclusion}
This paper proposed a novel approach for medical image synthesis using a generative model optimized by embedding a genetic algorithm, referred to as GAN-GA. The proposed model was tested on a small new medical image dataset, Acute Lymphoblastic Leukemia, and demonstrated enhanced generating performance and training stability compared to the baseline model, InfoGAN. The GAN-GA model also improved image fidelity and diversity while preserving distinctive features, which is crucial for accurate image interpretation. The experimental results showed that the proposed optimized GAN-GA enhanced the Frechet Inception Distance scores by about 6.8\%, especially in earlier training epochs. Integrating genetic algorithms with generative models proved to be an effective solution for medical image synthesis, producing a more exact outcome than traditional generative methods. Future research could explore applying the proposed GAN-GA model in other medical imaging domains and evaluate its performance against other state-of-the-art models.

\bibliographystyle{splncs04}

\end{document}